# On the anomalous low-resistance state and exceptional Hall component in hard-magnetic Weyl nanoflakes

Qingqi Zeng[1,†], Gangxu Gu[1,†], Gang Shi[1], Jianlei Shen[1], Bei Ding[1], Shu Zhang[2], Xuekui Xi[1], Claudia Felser[3], Yongqing Li[1,4], Enke Liu[1,4,*]

1. Beijing National Laboratory for Condensed Matter Physics, Institute of Physics, Chinese Academy of Sciences, Beijing 100190, China
2. Department of Physics and Astronomy, University of California, Los Angeles, California 90095, USA
3. Max Planck Institute for Chemical Physics of Solids, Dresden D-01187, Germany
4. Songshan Lake Materials Laboratory, Dongguan, Guangdong 523808, China

**Abstract**

Magnetic topological materials, which combine magnetism and topology, are expected to host emerging topological states and exotic quantum phenomena. In this study, with the aid of greatly enhanced coercive fields in high-quality nanoflakes of the magnetic Weyl semimetal $Co_3Sn_2S_2$, we investigate anomalous electronic transport properties that are difficult to reveal in bulk $Co_3Sn_2S_2$ or other magnetic materials. When the magnetization is antiparallel to the applied magnetic field, the low longitudinal resistance state occurs, which is in sharp contrast to the high resistance state for the parallel case. Meanwhile, an exceptional Hall component that can be up to three times larger than conventional anomalous Hall resistivity is also observed for transverse transport. These anomalous transport behaviors can be further understood by considering nonlinear magnetic textures and the chiral magnetic field associated with Weyl fermions, extending the longitudinal and transverse transport physics and providing novel degrees of freedom in the spintronic applications of emerging topological magnets.

**Introduction**

Recently, the first magnetic Weyl semimetal $Co_3Sn_2S_2$ was discovered[1-6], which has significantly improved the understanding of the interplay of magnetic order and topological physics after the experimental realization of the quantum anomalous Hall effect (QAHE)[7]. $Co_3Sn_2S_2$ is a Shandite compound with Weyl nodes at ~60 meV above the Fermi level and shows semi-metallicity with small Fermi surfaces and low carrier concentration. Transport properties are dominated by the electrons near Fermi energy. As the topological features in this system are sufficient close to the Fermi energy and there are few trivial bands, it is promising to observe the transport behavior dominated by topological band structures. In $Co_3Sn_2S_2$ bulk single crystals, large intrinsic anomalous Hall conductivity and anomalous Hall angle have been detected[1, 2]. In the two-dimensional (2D) limit, theoretical studies have predicted that the desired high-temperature QAHE could be realized[8-10], which has been indicated experimentally by the observation of chiral edge states in a recent STM measurement[11]. In addition, regardless of the specific types of magnetic domain walls, there may exist a large magnetoresistance (MR) that can be retained against the finite strength of the disorder in $Co_3Sn_2S_2$[10].

The application prospects of $Co_3Sn_2S_2$ make it a promising candidate for practical applications of topological materials in spintronic or advanced electronic devices. For topology-related spintronic applications and the realization of QAHE, the growth and physical study of $Co_3Sn_2S_2$ films are highly desired. Epitaxial polycrystalline $Co_3Sn_2S_2$ films can be grown via molecular beam epitaxy and co-sputtering deposition methods[12, 13]. Moreover, reports on $Co_3Sn_2S_2$ nanoflakes grown via the chemical vapor transport (CVT) method exist[14, 15]. Out-of-plane ferromagnetic order and large anomalous Hall effects were observed. These studies offer important indications for further studies of low-dimensional magnetic Weyl semimetals. In addition to studies on possible QAHE and its practical applications, the unexplored properties of this magnetic Weyl low-dimensional system are important. Anomalous transport properties in $Co_3Sn_2S_2$ nanoflakes have been observed in our previous work[14]. However, these intriguing behaviors still need to be studied and analyzed in detail.

In this work, anomalous transport properties are studied in high-quality $Co_3Sn_2S_2$ single-crystalline nanoflakes with 30–100-nm thickness. The system clearly enters a lower electrical resistance state once the external magnetic field is antiparallel to the magnetization. Furthermore, an exceptional Hall component besides the normal and anomalous Hall effects is observed concurrently. Possible magnetic textures induced by an antiparallel magnetic field are possibly responsible for the anomalous transport behaviors, as the interaction between Weyl electrons and magnetic textures can result in local states and conductive modes[16-20]. This study will attract interest in the interplay of magnetic order and topological transport behavior and offer a platform for magnetic topological material-based applications.

**Materials and methods**

Single-crystalline $Co_3Sn_2S_2$ nanoflakes were grown via CVT using $Co_3Sn_2S_2$ polycrystalline powder as the precursor. The morphology and thickness were investigated using atomic force microscopy. Hall bar pattern with a long edge along the *a* axis of $Co_3Sn_2S_2$ for out-of-plane transport measurement was fabricated using microfabrication technology (Supplementary Note 1). Transport data were measured using a Janis system and collected by employing lock-in amplifier technology. Symmetric and antisymmetric processes were performed on the longitudinal and transverse resistivity, respectively (details in Supplementary Note 3).

**Results and discussion**

The morphology and thickness of a single-crystalline nanoflake are shown in Figure 1(a) (more details are given in Supplementary Note 1). Samples of thicknesses 30, 41, 71, and 94 nm were studied in this work. Figure 1(b) shows one of the Hall bar patterns for transport measurements. The Curie temperatures of the two selected samples were determined as 181 K through the kink temperature in the temperature-dependent resistivity curve (Figure S3). The single-crystalline $Co_3Sn_2S_2$ nanoflakes being studied show consistent basic physical properties with bulk single crystals[1, 21] (Supplementary Note 4). Furthermore, the nanoflakes show high quality with a residual resistivity ratio of ~20, an anomalous Hall conductivity of ~1000 $\Omega^{-1}$ m$^{-1}$, and a large coercive field as high as 5.5 T.

Figures 1(c–j) show the magnetic field dependence of the longitudinal and transverse resistivity measured at various temperatures for the 30-nm-thick sample (data of other samples are shown in Supplementary Note 4). The longitudinal data (Figures 1(c–f)) show positive parabolic-like MR at all measured temperatures. Apart from the ordinary MR, hysteresis behavior exists where the data with decreasing and increasing fields are not coincident between zero and coercive fields ($\mu_0H_c$). Thus, the value of the longitudinal resistivity depends on the magnetization history, being lower (50 K) or higher (1.6 K) for the demagnetization compared with the magnetization process.

As shown in Figures 1(g–j), the Hall data exhibit hysteresis loop behavior like the shape of the magnetic field dependence of magnetization in a hard-ferromagnetic material. The anomalous Hall resistivity, in addition to the normal Hall effect, is detected. The loop behavior and the anomalous Hall effect like that of the bulk sample[1] suggest that the easy axis of $Co_3Sn_2S_2$ single-crystalline nanoflake is still along the $c$ axis. Clear nonlinear normal Hall curves are observed at low temperatures, reflecting that both electron and hole carriers contribute to the transport properties in $Co_3Sn_2S_2$ single-crystalline nanoflakes. We performed two-band[1, 22, 23] analyses on all measured samples. The carrier concentration of holes and electrons are both ~1 $\times 10^{20}$ cm$^{-3}$ in all studied samples (details in Supplementary Note 4), indicating a near compensation of carriers in $Co_3Sn_2S_2$ single-crystalline nanoflakes with the abovementioned different thicknesses. The abovementioned results reveal the consistent carrier properties of $Co_3Sn_2S_2$ single-crystalline nanoflake and bulk samples[1, 14, 15].

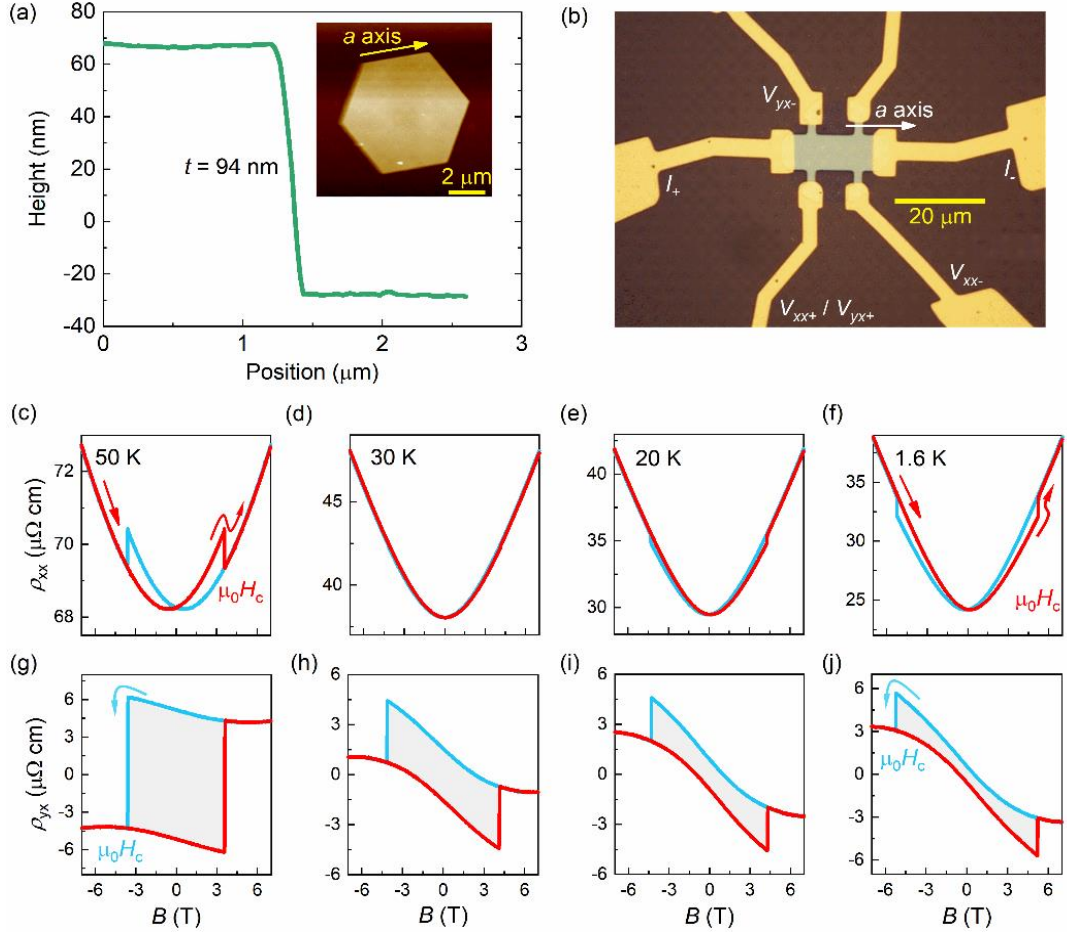

**Figure 1**. (a) Thickness and morphology (inset) of the 94-nm-thick sample determined using atomic force microscopy. (b) Optical image of the Hall bar shape 41-nm-thick sample. A type of electrode connection path is denoted. Current is applied along the *a* axis (indicated by a white arrow) of $Co_3Sn_2S_2$. (c–f) Longitudinal resistivity measured at 50, 30, 20, and 1.6 K for the 30-nm-thick sample. The arrows indicate the direction of field variation. (g–j) Hall data measured simultaneously with the longitudinal resistivity in (c–j).

The additional MR peak or dip emerging in the background of the parabolic-like MR (the curves without magnetization reversal, i.e., from ±9 T to zero fields) was extracted and denoted as $\Delta\rho_{xx}$. At 50 K, when the field decreased from ±9 T to zero and then increased in the opposite direction, a linear field-dependent positive $\Delta\rho_{xx}$ appeared immediately and became zero instantaneously after the magnetization reversal (Figure 2(a)). However, a negative linear $\Delta\rho_{xx}$ was unexpectedly observed at 1.6 K when the applied field approached the coercive field (Figure 2(b)).

We further analyzed the emerging additional MR when magnetization was antiparallel to the applied magnetic field. In a widely accepted case, a system will be more magnetically ordered when the magnetic field is much higher than $\mu_0 H_c$. Between zero and coercive fields, the magnetic field gradually turns the antiparallel moments into parallel moments. Thus, the antiparallel moments become more disordered as the field increases before the magnetization reversal. Hence, the spin-dependent scattering[7, 24, 25] would be strong, increasing the longitudinal resistivity. Therefore, positive $\Delta\rho_{xx}$, like the behavior in Figure 2(a), is normally observed in magnetic systems[7, 26]. The difference from the common case is that the positive $\Delta\rho_{xx}$ in the studied sample appears immediately when the antiparallel magnetic field starts increasing. More widely observed positive MR peaks appear only near the coercive field[27-29]. The linear magnetic field dependence is also a unique and unexplained phenomenon.

Interestingly, a contrary situation emerges when the system goes to low temperatures. The longitudinal resistivity in the case when the applied field is antiparallel to the magnetization (antiparallel configuration) is even lower than that of the parallel case (Figure 2(b)), i.e., $\Delta\rho_{xx}$ is negative. This behavior has been previously observed in our 180-nm-thick $Co_3Sn_2S_2$ single-crystalline nanoflakes[14] and can also be recognized in polycrystalline films (although the spin-related positive MR is mixed in)[13]. Evidently, the trend is a universal behavior in this low-dimensional Weyl system. Nevertheless, negative $\Delta\rho_{xx}$ has scarcely been observed among common magnetic materials or magnetic functional films in the past. A distinct mechanism resulting in a low-resistance state against the antiparallel external field may emerge in this magnetic topological system.

In magnetic information storage technology, the giant MR effect can be produced by switching the parallel and antiparallel directions of ferromagnetic and antiferromagnetic layers in magnetic tunnel junctions, which use different electrical resistance states between parallel and antiparallel cases[30-32]. The anomalous negative MR effect observed in this magnetic nanoflake may provide a novel degree of freedom for spintronic applications.

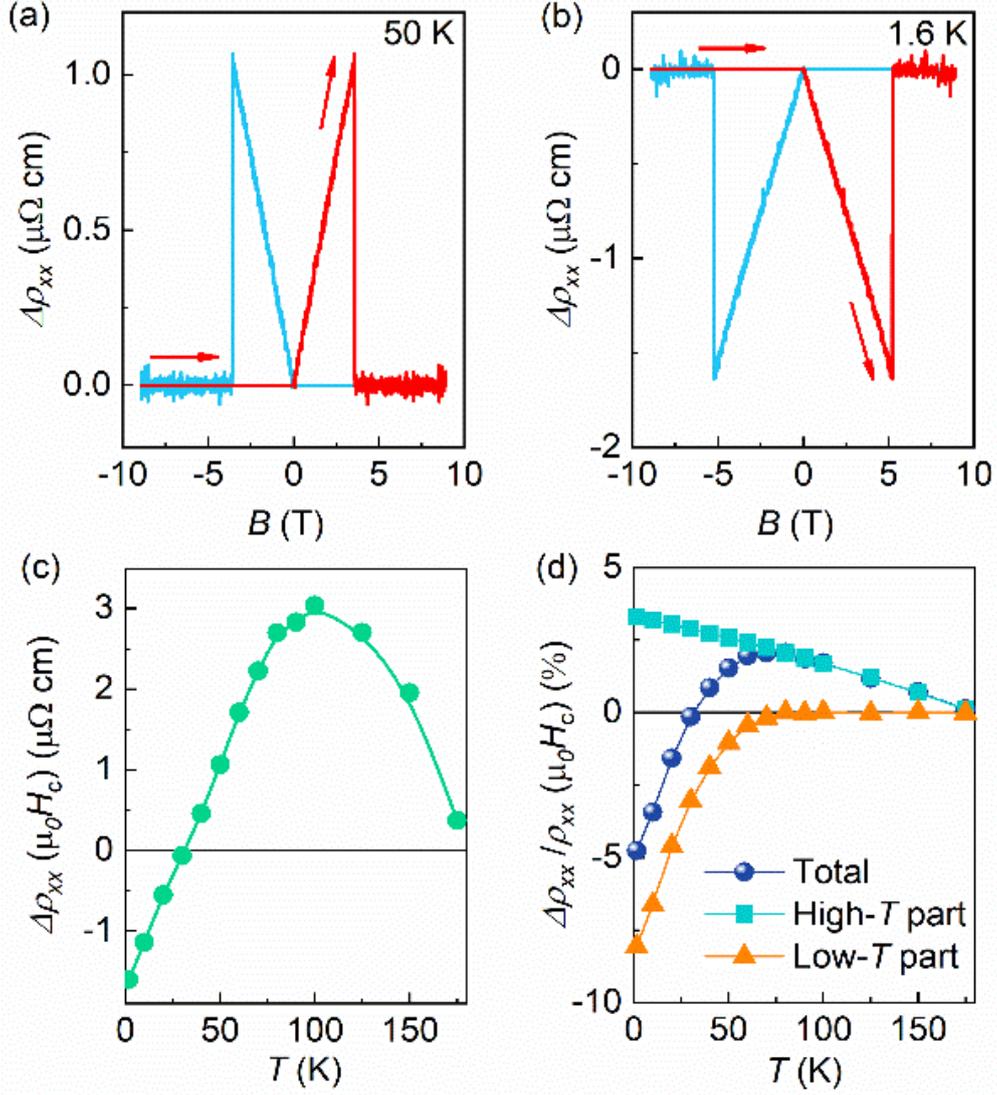

**Figure 2.** (a–b) Parabolic-like MR-subtracted magnetic field dependence of the longitudinal resistivity ($\Delta\rho_{xx}$) at 50 and 1.6 K for the 30-nm-thick sample. The arrows indicate the direction of field variation. (c) $\rho_{xx}$ difference before and after the jump at $\mu_0 H_c$ ($\Delta\rho_{xx}(\mu_0 H_c)$). (d) Ratio of $\Delta\rho_{xx}$ to $\rho_{xx}$ on the parabolic-like MR background at $\mu_0 H_c$ (dark blue dots) and separate contributions (green and orange symbols).

Figures 2(c–d) show $\Delta\rho_{xx}$ at $\mu_0H_c$ ($\Delta\rho_{xx}(\mu_0H_c)$) and the $\Delta\rho_{xx}(\mu_0H_c)$-to-$\rho_{xx}(\mu_0H_c)$ ratio on the background of the parabolic-like MR (dark blue dots). Above and below 30 K, $\Delta\rho_{xx}(\mu_0H_c)$ is positive and negative, respectively. As a kink of $\Delta\rho_{xx}(\mu_0H_c)$ exists around 100 K, we considered the existence of competing components leading to different field dependences of longitudinal resistivity and their dominant position variation with temperature. For simplicity, two factors are currently assumed in this system. To separate these two parts, we assume that one part, called the low-temperature part (low-$T$ part), occurs only below the kink temperature in $\Delta\rho_{xx}(\mu_0H_c)$. The other part is the high-temperature part (high-$T$ part), which is fitted using the data just above 100 K. By adopting the form of $y = A\,e^{-x/B} + y_0$, the experimental data (dark blue dots) can be well fitted (green square symbols) above 80 K (see Figure S6 for details). The low-$T$ part is then extracted by subtracting the high-$T$ part from the experimental results, denoted by orange triangle symbols in Figure 2(d). The current straightforward but effective fitting indicates that the high-$T$ part contributes to longitudinal resistivity continuously below the Curie temperature. By contrast, the low-$T$ part appears only below 70 K. Here, we stress that the low-resistance state can be observed directly from the original MR data without separation (Figures 1(e) and 1(f)). Thus, the current separation will not affect the main results.

We next analyze the transverse transport behavior in $Co_3Sn_2S_2$ single-crystalline nanoflakes. Conventional magnetic materials usually show both normal and anomalous Hall effects. For a hard-magnetic material with coercivity and sharp magnetization reversal, the typical Hall curve is shown in the upper panel of Figure 3(a). We denote the Hall resistivity curve from +9 T to −9 T as $\rho_{yx}^-$. Similarly, the curve from −9 T to +9 T is denoted as $\rho_{yx}^+$. Evidently, $\rho_{yx}^- - \rho_{yx}^+$ maintains a stable plateau (two times as anomalous Hall resistivity ($\rho_{AH}$)) within the coercive field (the lower panel of Figure 3(a)). For $Co_3Sn_2S_2$ nanoflakes, the difference between the Hall data with different directions of the magnetic field variation is shown in Figure 3(b). In sharp contrast, an increasing $\rho_{yx}^- - \rho_{yx}^+$ effect exists with increasing applied fields up to the coercive field, which is quite different from the conventional case.

The corresponding Hall loop, subtracting the normal Hall component, is depicted in Figure 3(c). In a conventional hard-magnetic material, the normal Hall-subtracted Hall curve should be a square loop[7], indicated by the red-dashed lines in Figure 3(c). However, unexpected Hall contribution (highlighted by light-blue-colored filling, denoted as exceptional Hall resistivity $\Delta\rho_{yx}$ in this text) exists except normal and anomalous Hall components in the current system.

Figure 3(d) shows the $\Delta\rho_{yx}$ at different temperatures in the negative field region. The $\Delta\rho_{yx}$ appears from the zero-field and increases noticeably with increasing antiparallel field up to the coercivity. Once the magnetization and applied field are parallel to each other after the magnetization reversal at the coercive field, the exceptional Hall resistivity suddenly drops to zero. Compared with the bulk case having a small coercive field (~0.5 T)[1], the enhanced coercivity in $Co_3Sn_2S_2$ nanoflake makes it possible to clearly observe $\Delta\rho_{yx}$ in a wide antiparallel field range. By further adopting the function form of $\Delta\rho_{yx} = a \cdot B^b$, $\Delta\rho_{yx}$ is found to evolve from a ~$B^{2.5}$ law at 50 K to a stable ~$B^{1.8}$ law below 30 K (Figure 3(e)).

Evidently, $\Delta\rho_{yx}$ emerges around 70 K and becomes more notable at lower temperatures. A large value of 1.5 μΩ cm is observed at 1.6 K. Because the low-$T$ negative MR (Figure 2(d)) also appears around 70 K, we conclude that there should be connections between anomalous longitudinal and transverse transport behaviors.

Figure 3(f) shows $\Delta\rho_{yx}$ at $\mu_0H_c$ and its ratio to $\rho_{AH}$. Evidently, the exceptional Hall component can be more than 2.5 times larger than $\rho_{AH}$, indicating a notable effect driven by a distinct mechanism in $Co_3Sn_2S_2$ nanoflakes at low temperatures.

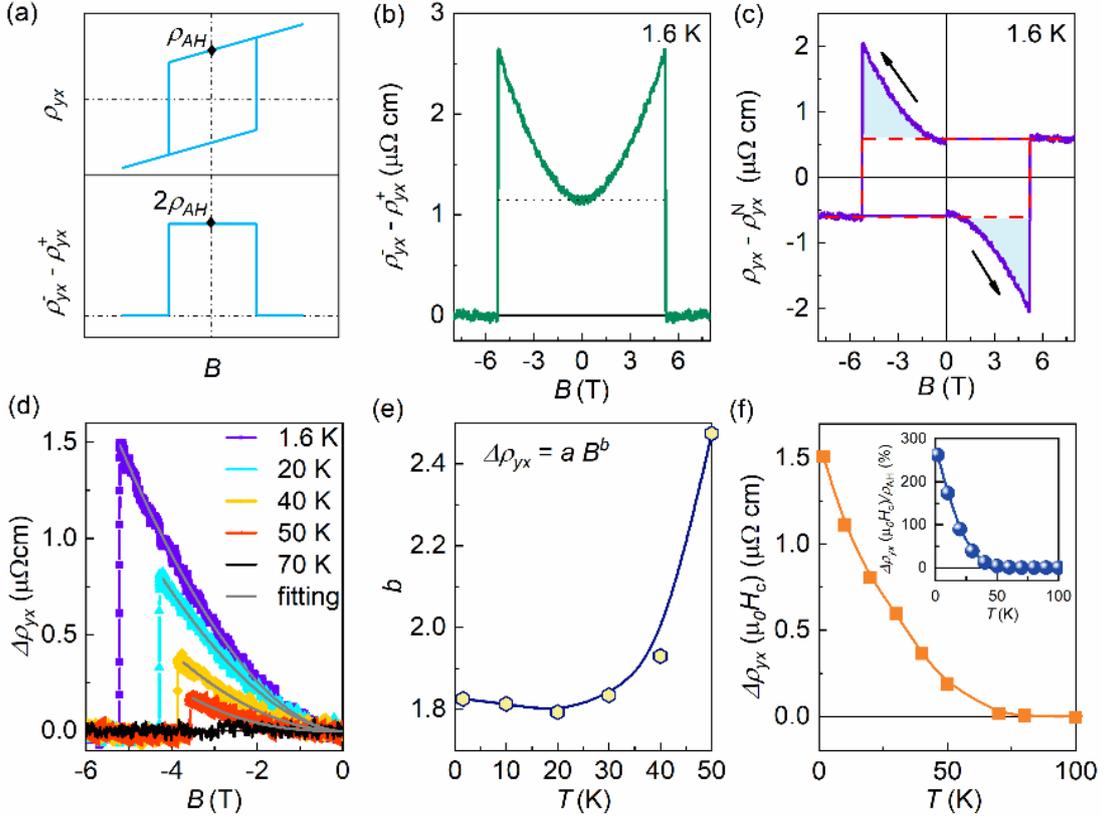

**Figure 3.** Transverse resistivity analyses of the 30-nm-thick sample. (a) Sketches of the Hall loop in a conventional hard-magnetic material (upper panel). The lower panel shows the value of $\rho_{yx}^- - \rho_{yx}^+$, where the Hall resistivity curve from +9 T (−9 T) to −9 T (+9 T) is denoted as $\rho_{yx}^-$ ($\rho_{yx}^+$). (b) $\rho_{yx}^- - \rho_{yx}^+$ in the sample being studied. (c) Normal Hall-subtracted transverse resistivity. (d) Exceptional Hall resistivity, denoted as $\Delta\rho_{yx}$. The gray lines are fitting results obtained by adopting $\Delta\rho_{yx} = a \cdot B^b$. (e) Temperature dependence of fitting parameter $b$. (f) $\Delta\rho_{yx}$ at the coercive field. The inset shows the $\Delta\rho_{yx}(\mu_0 H_c)$-to-$\rho_{AH}$ ratio.

Figures 4(a–b) summarize the anomalous transport behaviors of four different-thick samples. The Hall conductivity analysis can be found in Supplementary Note 6. All samples show similar temperature-dependent transport

behaviors, exhibiting anomalous negative MR and exceptional Hall resistivity below 70 K. According to the data, no obvious thickness dependence exists for these anomalous transport behaviors in the current thickness range. In addition, the 41-nm-thick sample shows smaller negative MR and exceptional Hall values compared with other samples. Further, it shows lower mobility (Figure S5(b)), possibly attributed to the deviation of sample quality.

Next, we analyze the underlying origin of the anomalous transport behaviors. The longitudinal resistance states for a certain magnetic field value are different during field-decreasing and -increasing processes. Thus, the anomalous transport behaviors could not result from the samples' basic properties (e.g., carrier mobility), which may vary with the magnetic field. The two-band analysis (Figure S5(c)) also indicates additional components apart from the normal and anomalous Hall conductivities, emerging only under an antiparallel magnetic field. Considering the coexistence of negative MR and exceptional Hall effect, these two effects may simultaneously originate from a specific physical state that emerges under an antiparallel magnetic field at low temperatures. It has been reported that nontrivial topological magnetic textures may emerge in a magnetic Weyl system[19, 33-35]. Once the magnetic textures (even without topological property) are formed, they will interact with the Weyl conduction electrons and affect the transport behavior in a system through an emergent chiral magnetic field[16-20]. The fictitious magnetic field may be reminiscent of the real-space Berry curvature in conventional materials with trivial band structures. However, the chiral magnetic field shows a richer degree of freedom as the related chiral gauge potential is closely linked to the Weyl nature and depends on the direction of the background magnetization itself[19, 20]. Theoretical calculations show that there could be bound states and potential current, which are highly localized around a magnetic domain wall or texture[16, 17, 20]. In experiments, $Co_3Sn_2S_2$ bulk samples indeed showed other possible magnetic interactions besides the out-of-plane ferromagnetic interaction, and thus, nonuniform magnetization behavior below the Curie temperature[1, 21, 36, 37]. A round-shaped domain structure has also been observed in the $Co_3Sn_2S_2$ system[38]. Based on the above theoretical and experimental reports, a promising possibility responsible for the anomalous negative MR and exceptional Hall effect in $Co_3Sn_2S_2$ nanoflakes is the noncollinear magnetic structure. Based on our experiment results, we consider it the magnetic texture induced by the antiparallel magnetic field.

Figures 4(c–d) show schematic illustrations of the magnetic structure when an external magnetic field is antiparallel to the magnetization. As the magnetization reversal process is sharp, we consider the antiparallel magnetic domains formed close to the coercivity sweeping out the entire sample immediately. Hence, only a slight tilt of moments occurs when the applied field is less than the coercivity. Moments may tilt randomly owing to the antiparallel magnetic field perturbation at temperatures higher than 70 K (Figure 4(c)), while the ordered magnetic textures (sketched as red patterns in Figure 4(d)) may emerge at low temperatures due to the competition of ferromagnetic and other magnetic interactions.

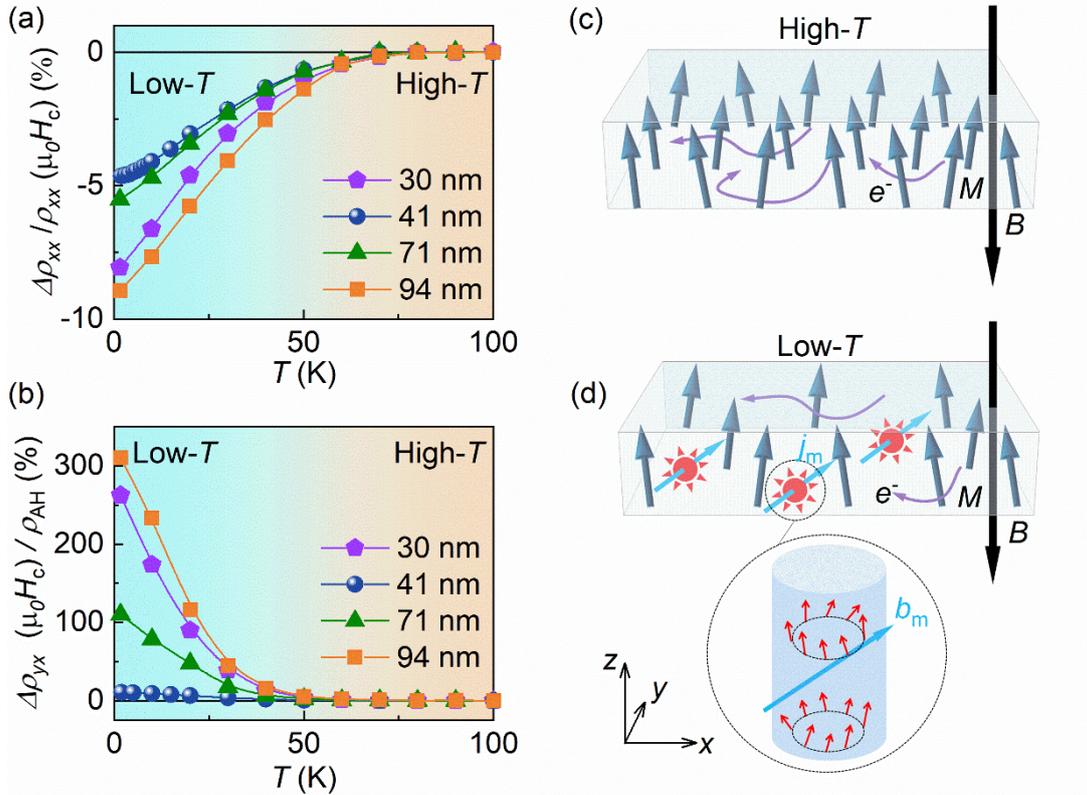

**Figure 4.** (a) $\Delta\rho_{xx}(\mu_0 H_c)$-to-$\rho_{xx}(\mu_0 H_c)$ ratio of the low-$T$ part. (b) $\Delta\rho_{yx}(\mu_0 H_c)$-to-$\rho_{AH}$

ratio. Sketches of magnetic moments (dark blue arrows) and locally ordered magnetic textures (red patterns) at high (c) and low (d) temperatures. The purple arrows indicate the scattered electrons. Blue arrows denote possible directions of the localized chiral magnetic field ($b_m$) and current ($j_m$).

Once the abovementioned potential magnetic textures are induced by the antiparallel magnetic field, a Weyl-fermion-correlated chiral gauge field $\vec{b}_m(\vec{r}) = \nabla \times \frac{J_{ex}}{v_F} M(\vec{r})$ [16-20] (where $J_{ex}$ is the exchange interaction strength between localized magnetic moments and Weyl electrons, and $v_F$ is the Fermi velocity) emerges around the magnetic textures. Furthermore, this localized chiral gauge field may lead to bound states near the textures[16, 17, 20]. When an electric field is applied to the sample, potential bound states can also offer emergent conductive modes in this system. As anomalous transport behaviors exist in both longitudinal and transverse channels in Co$_3$Sn$_2$S$_2$ nanoflakes, a varying $M(\vec{r})$ with a complicated order (not varied within parallel planes, schematic sketched in Figure 4(d)) may occur, leading to $j_m$ with both longitudinal (corresponding to low-resistance state) and transverse (corresponding to exceptional Hall effect) components. The specific structure of the possible magnetic textures in Co$_3$Sn$_2$S$_2$ single-crystalline nanoflakes and the confirmation of its influence on the transport behaviors require further study.

**Conclusions**

Herein, an anomalous low electrical resistance state and exceptional Hall component are reported in high-quality single-crystalline nanoflakes of magnetic Weyl semimetal Co$_3$Sn$_2$S$_2$ once the magnetization is antiparallel to external magnetic fields. Possible antiparallel field-induced magnetic textures and Weyl-fermion-associated chiral magnetic fields are expected to account for the decrease and increase in the longitudinal and transverse resistivities, respectively, through the possible additional conductive modes. Our report will draw more interest on the interplay of magnetism and topological transport phenomena, offering new elementary states for the applications of magnetic topological materials.

**Supplemental material**

Supplemental material is available online or from the author.

**Acknowledgments**

Qingqi Zeng and Gangxu Gu contributed equally to this work.

This work was supported by the National Natural Science Foundation of China (Nos. 52088101 and 11974394), National Key R&D Program of China (No. 2019YFA0704900), Beijing Natural Science Foundation (No. Z190009), the Strategic Priority Research Program (B) of the Chinese Academy of Sciences (CAS) (No. XDB33000000), the Scientific Instrument Developing Project of CAS (No. ZDKYYQ20210003), Users with Excellence Program of Hefei Science Center CAS (No. 2019HSC-UE009), and the Youth Innovation Promotion Association of Chinese Academy of Sciences (No. 2013002). S. Z. is supported by NSF under Grant No. DMR-1742928.